# Programmable metasurface-based RF chain-free 8PSK wireless transmitter

Wankai Tang, Jun Yan Dai, Mingzheng Chen, Xiang Li, Qiang Cheng, Shi Jin✉, Kai-Kit Wong and Tie Jun Cui

In this Letter, a wireless transmitter using the new architecture of programmable metasurface is presented. The proposed transmitter does not require any filter, nor wideband mixer or wideband power amplifier, thereby making it a promising hardware architecture for cost-effective wireless communications systems in the future. Using experimental results, the authors demonstrate that a programmable metasurface-based 8-phase shift-keying (8PSK) transmitter with 8 × 32 phase adjustable unit cells can achieve 6.144 Mbps data rate over the air at 4.25 GHz with a comparable bit error rate performance as the conventional approach without channel coding, but with less hardware complexity.

*Introduction:* The fifth-generation wireless communication will soon be delivering the needed ultra-high data rates, for which massive multiple-input multiple-output (MIMO) antennas and millimetre-wave (mm-wave) are the two key technologies. However, the need for large number of high-performance RF chains means that there is an increase in the hardware complexity, leading to high cost and power consumption [1]. Although solutions such as analogue beamforming [2] and hybrid beamforming [3] have been developed in the literature, to reduce the number of RF chains, they do not fundamentally eliminate the hardware constraints. It is therefore of great significance to develop a more flexible hardware architecture for future wireless communications.

On the other hand, recent advances in programmable metasurfaces have attracted much attention due to their superior capability in controlling the amplitude or phase of electromagnetic waves in real time [4]. Manipulation of reflected or transmitted electromagnetic waves can be achieved by electrically changing the reflection or transmission coefficient distributions on the metasurfaces [5], effectively realising modulation of electromagnetic waves, which has appeal in applications of wireless communications. However, using metasurface in wireless communication is still in its infancy. A recent study, in [6] where a metasurface-based binary frequency shift-keying transmitter was presented, in which the two signal frequencies representing '0' and '1' respectively were synthesised directly via a programmable metasurface. However, only a low data rate of 78.125 kbps was reported.

In this Letter, we take advantage of a programmable metasurface that achieves 360° phase response coverage of reflected electromagnetic wave to develop a novel metasurface-based 8-phase shift-keying (8PSK) transmitter. The experimental results validate its feasibility by observing the demodulated constellation diagrams and measuring the bit error rate (BER) performance with an extraordinary 6.144 Mbps data rate over the air.

*Design:* As shown in Fig. 1, the unit cell of metasurface can be regarded as a resonant tank with the load impedance $Z_l$ and the characteristic impedance of the air $Z_0$. Also, $\Gamma$ in the figure represents the reflection coefficient, which describes how much of the incident electromagnetic wave is reflected by an impedance discontinuity in the transmission medium, and equals to the ratio of the complex amplitude of the reflected wave to that of the incident wave.

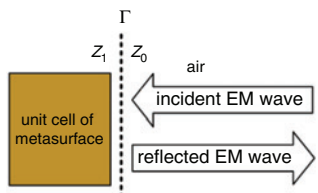

**Fig. 1** *Resonant tank model of the metasurface unit cell*

In particular, the reflection coefficient of the unit cell of the metasurface can be written as

$$\Gamma = \frac{Z_l - Z_0}{Z_l + Z_0},$$

where $Z_0$ is a constant with the value of 377 Ω. The desired $Z_l$ can be achieved by carefully designing the physical structure of the unit cell, thus realising the desired phase or amplitude response of the reflected electromagnetic wave. Note that conventionally, the reflection coefficient of the metasurface is fixed once the metasurface is designed and fabricated. Nevertheless, recent advances in programmable metasurface allow some tunable components in their unit cells such as varactor diodes, which make it possible to dynamically change $Z_l$ in real time and manipulate the phase or amplitude of the reflected electromagnetic wave in a controllable and programmable manner.

The programmable metasurface utilised in our transmitter has been elaborately designed to have a full 360° phase response and its reflectivity is about 85% at the working frequency of 4.25 GHz. That is to say, it can regulate the phase of an electromagnetic wave flexibly and efficiently. Fig. 2 depicts a fabricated sample of our programmable metasurface with a zoomed-in view of the unit cell. There are 8 × 32 unit cells on the entire surface and in each unit cell, five planar patterned metallic patches are printed on the top surface of the F4B substrate with a dielectric constant of 2.65 and a loss tangent of 0.001, with two varactor diodes and two little capacitors loaded between them. The size of the unit cell is $12 \times 12 \times 5$ mm$^3$, which is approximately $0.17 \times 0.17 \times 0.07$ $\lambda^3$ at 4.25 GHz. This well-designed structure greatly improves the unit cell in [6], which consists of only two metallic patches and one varactor diode so that only one resonant pole can be obtained and a 360° phase regulation of the reflected wave is not achievable. By contrast, the two varactor diodes in the new design can form two resonant poles, allowing a theoretical 720° phase shift and ensuring a practical 360° full phase manipulation. Also, the capacitance of the unit cell is dominated by the varactor diode, which indicates that the phase response of the metasurface can be tuned by the bias voltages of varactor diodes.

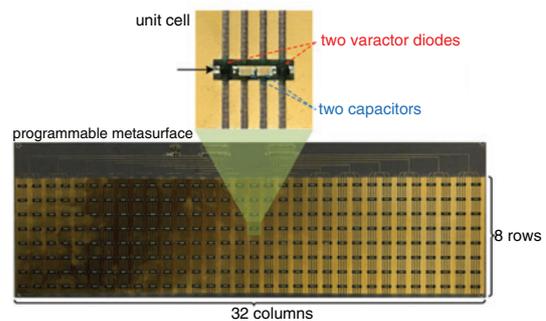

**Fig. 2** *Fabricated sample of the programmable metasurface with a zoomed-in view of the unit cell that was used in the experiment*

The relationship between the bias voltage of the varactor diode and the phase of the reflected electromagnetic wave of each unit cell is shown in Fig. 3. The bias voltage has an approximate linear relationship with the phase of the reflected wave and can achieve 360° phase modulation coverage. For example, eight bias voltages corresponding to the eight red points shown in Fig. 3 can be applied to achieve 8PSK modulation. Each red point corresponds to a specific bias voltage and the desired phase state of the reflected wave for 8PSK modulation.

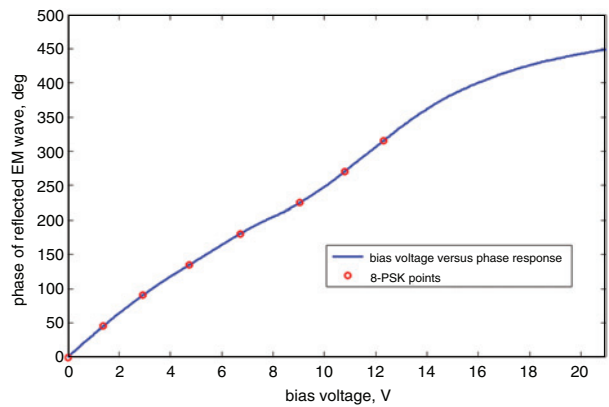

**Fig. 3** *Relationship between the bias voltage and the phase of the reflected electromagnetic wave*

Fig. 4a shows the hardware architecture of the proposed transmitter based on the mentioned programmable metasurface. Unlike the conventional transmitter, in which a local oscillator provides the carrier signal to the mixer over a transmission line [7], the metasurface-based transmitter obtains the carrier signal from a feed antenna through the air. The air-fed carrier signal makes the same carrier signal naturally realised between each transmitting unit cell of the metasurface. The digital-to-analogue converters (DAC) convert the digital baseband sequences into analogue voltage signals to control each unit cell of the metasurface, thus achieving phase modulation of the incident single-tone carrier signal and generating the reflected RF signal with modulated information.

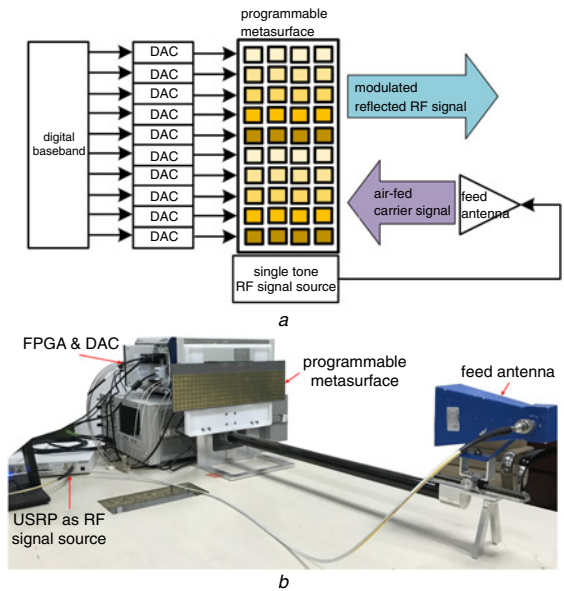

**Fig. 4** *Proposed programmable metasurface-based transmitter*
*a* Hardware architecture
*b* Hardware realisation

Fig. 4b illustrates the actual hardware realisation of the proposed metasurface-based transmitter, which is composed of the programmable metasurface, the feed antenna, the universal software radio peripheral (USRP, a commercial software-defined radio (SDR) transceiver) as RF signal source, the field programmable gate array (FPGA) as baseband module and the DAC module. The FPGA generates the corresponding 8PSK baseband digital sequence. The DAC converts the baseband sequence into the varying bias voltage signal and provides it to the metasurface. The USRP produces a 4.25 GHz single-tone carrier signal, which is fed to the metasurface through the feed antenna.

*Test-bed setup:* The above metasurface-based transmitter comprises 256 unit cells. Since each unit cell can be controlled independently by a separate DAC in principle, 256 channels of data can be radiated simultaneously. That is to say, the proposed transmitter can generate complex RF signals [4], which as a result enables complex signal processing methods such as beamforming and space–time modulation for MIMO antenna technologies. In addition, Fig. 5 shows a photo of the testbed of the proposed programmable metasurface-based transmitter. In this Letter, we aim to investigate the feasibility of using this new architecture for realising 8PSK wireless communication. In this case, all the unit cells on the programmable metasurface are controlled by the same bias voltage, which contains the phase modulation information of only one data stream. Additionally, because the air-fed carrier signal can be regarded as a plane wave approximately, the reflected electromagnetic waves of each unit cell are the same and uniformly radiated to all directions ahead. The radiation of unit cells will superimpose with each other in the space, thus forming a single beam [4, 5].

In the test, the metasurface-based transmitter constantly sends an 8PSK RF signal, in which each frame contains one sync subframe, one pilot subframe and nine data subframes. The receiver USRP performs baseband signal processing and recovers the transmitted data bits. This testbed has two operation modes: metasurface-based mode and conventional mode. In the metasurface-based mode, 8PSK modulation is implemented by the metasurface as described above. In the conventional mode, the USRP transmitter provides the 8PSK modulated RF signal directly instead of a carrier signal, while the control voltage of the metasurface remains fixed, i.e. the metasurface only acts as a reflector. Therefore, we can compare the performance between the programmable metasurface-based transmitter and the conventional one.

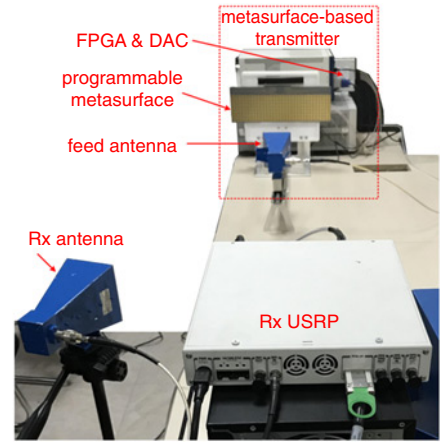

**Fig. 5** *Our testbed of the proposed programmable metasurface-based transmitter for 8PSK wireless communication*

*Measurement results:* A series of measurements are conducted indoors with a distance of three metres from the transmitter to the receiver in the above testbed. Fig. 6 reveals the relationship between the BER and the transmission rate when using the metasurface based or the conventional transmitter with −22 dBm transmit power. The BER performances of both approaches decrease as the transmission rate becomes higher, but it gets worse faster at high transmission rates for the metasurface-based transmitter due to the capacitance-induced charge and discharge effects in the unit cells. With a 6.144 Mbps transmission rate, the BER with the metasurface-based transmitter is $3 \times 10^{-3}$, which will result in reasonable performance using forward error correction in practice.

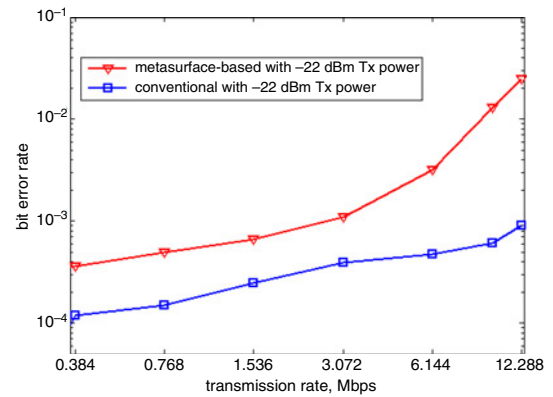

**Fig. 6** *BER versus transmission rate*

The signal-to-noise ratio (SNR) and the corresponding BER at the receiver are measured and they are plotted against each other in Fig. 7. As can be observed, the BER versus SNR curve for the metasurface-based transmitter almost coincides that of the conventional SDR-based transmitter. Such result is expected because they use the same wireless frame structure, same transmission rate, same baseband demodulation algorithm and go through the same wireless communication channel in the comparative experiment. This is why the same SNR value corresponds to the same BER performance here, regardless of whether it is the conventional or metasurface-based transmitter that is used.

To compare the performance difference between the two schemes further, it is necessary to consider the relationship between the transmission power and the BER, which will reveal how well the transmitter converts the transmission power into the effective band energy.

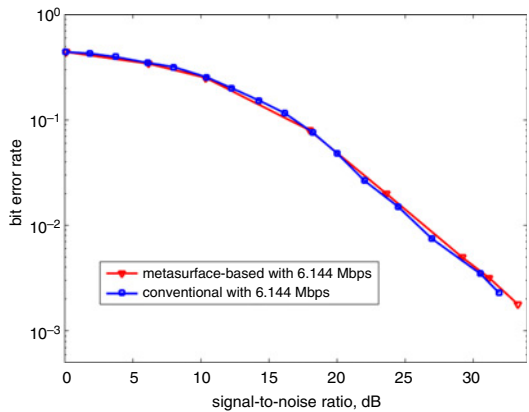

**Fig. 7** *BER versus SNR*

On the basis of the aforementioned discussion, the BER performance with different transmission power is measured, as shown in Fig. 8 and we compare the BER performance between the programmable metasurface-based transmitter and the conventional transmitter. As we can see, with 6 dB increase of the transmission power, the same BER performance can be obtained as the conventional one. Given the advantages of low hardware cost and simple structure, such result is encouraging for the proposed metasurface-based transmitter, rendering it an attractive new hardware architecture for future wireless communications.

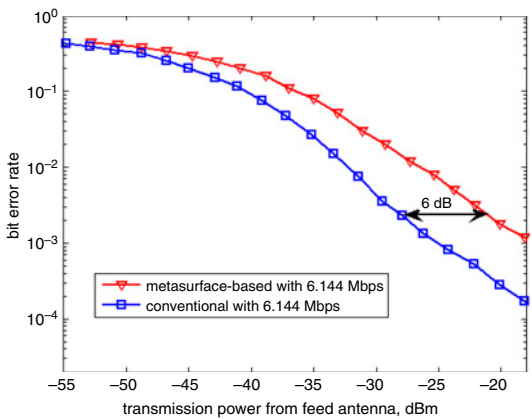

**Fig. 8** *BER versus transmission power*

The measured constellation diagrams under different transmission power for the proposed metasurface-based transmitter with 6.144 Mbps transmission rate are illustrated in Fig. 9. As we can see, if the transmission power increases, the demodulated constellation points will become more concentrated, which indicates better BER performance.

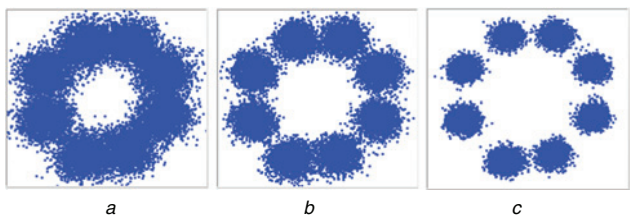

**Fig. 9** *Constellation diagrams under different transmission power*
*a* – 40 dBm
*b* – 30 dBm
*c* – 20 dBm

Finally, Table 1 compares the hardware requirements of the metasurface-based and conventional transmitters with 256 channels. Apparently, each channel of the conventional transmitter contains one power amplifier (PA), two mixers and at least two filters [7]. However, the metasurface-based transmitter only requires one narrow band PA to control the power of the air-fed single-tone carrier signal without the need for mixers and filters, no matter how many channels it has. This is a huge advantage of the proposed architecture compared to the conventional one. Therefore, although the performance of the metasurface-based transmitter is slightly inferior to the conventional one as shown in Figs. 6 and 8, it is still an attractive cost-effective transmitter architecture, especially for future communication systems with a massive number of RF chains.

**Table 1:** Number of RF components contained in the metasurface-based transmitter and conventional transmitter with 256 channels

| Name | Metasurface-based transmitter with 256 channels | Conventional transmitter with 256 channels |
| --- | --- | --- |
| Number of PAs | 1 | 256 |
| Number of mixers | 0 | 512 |
| Number of filters | 0 | 512 |

*Conclusion:* A transmitter totally based on the programmable metasurface has been demonstrated. Real-time transmission over the air with a 6.144 Mbps data rate under 8PSK modulation was achieved. This new architecture eliminates the need for mixers, filters and wideband PAs while providing comparable BER performance as the conventional architecture, thus leading to a cost-effective hardware architecture with great potential for future wireless communications.

*Acknowledgments:* This work was supported in part by the National Science Foundation (NSFC) for Distinguished Young Scholars of China with grant no. 61625106 and the National Natural Science Foundation of China under grant no. 61531011.

This is an open access article published by the IET under the Creative Commons Attribution License (http://creativecommons.org/licenses/by/3.0/)
Submitted: *31 January 2019* E-first: *12 March 2019*
doi: 10.1049/el.2019.0400
One or more of the Figures in this Letter are available in colour online.

Wankai Tang, Jun Yan Dai, Mingzheng Chen, Xiang Li, Qiang Cheng, Shi Jin and Tie Jun Cui (*School of Information Science and Engineering, Southeast University, People's Republic of China*)

✉ E-mail: jinshi@seu.edu.cn

Kai-Kit Wong (*Department of Electronic and Electrical Engineering, University College London, United Kingdom*)